\documentstyle[prl,aps]{revtex}

\def\q1{{q^{-1}}}
\def\qq1{{q-q^{-1}}}

\def\dqq{{{\cal D}^{(q)}}}
\def\nq{{n_{i}}}

\begin{document}

\title{Generalized thermodynamics of $q$-deformed bosons and fermions}
\author{A. Lavagno $^{a,b}$ and P. Narayana Swamy $^c$}
\address{
$^a$ Dipartimento di Fisica, Politecnico di Torino, I-10129 Torino, Italy \\
$^b$ Istituto Nazionale di Fisica Nucleare, Sezione di Torino, I-10125, Italy \\
$^c$ Professor Emeritus,  Southern Illinois University,
Edwardsville, IL 62026, USA}
\maketitle

\begin{abstract}

We study the thermostatistics of $q$-deformed bosons and fermions 
obeying the symmetric ($q\leftrightarrow\q1$)  algebra and show that  it
can be  built on the 
formalism of $q$-calculus. The entire 
structure of thermodynamics is preserved if ordinary derivatives  
are replaced by an appropriate Jackson derivative. 
In this framework, we derive the most important thermodynamic functions 
describing  the $q$-boson and $q$-fermion ideal gases in the thermodynamic limit. 
We also investigate the semi-classical limit and the low temperature regime and demonstrate
that the nature of the $q$-deformation gives rise to  pure quantum statistical effects 
stronger than undeformed boson and fermion particles.  

\end{abstract}

\section{Introduction}

There have emerged two distinct methods in the literature for the purpose of introducing an intermediate stati\-sti\-cal behavior to describe a deformed physical system. The first method consists of modifying the partitioning of the microstates and thus the logarithm of the statistical weight $W$ of the many body system. Specifically this is accomplished by a deformation of the logarithm function in the Boltzmann entropy $S = \log_T W$,  by the introduction of the Tsallis logarithm \cite{tsallis}. This then introduces  nonextensivity in the statistical behavior. The second method is to deform the quantum algebra of the creation and annihilation operators, thus modifying the exchange factor between the permuted particles or states. The theory of $q$-oscillators is related to the theory of quantum groups originally introduced by Biedenharn and Macfarlane \cite{bie,mac}. 

The mathematical framework of $q$-oscillators is based on the 
$q$-calculus which is introduced via the Jackson derivative (JD) \cite{jack}. We thus expect such a 
$q$-calculus to play an important r\^ole in the thermostatistics of $q$-oscillators. 
Indeed, it has been shown  \cite{pre2000} that an internally self-consistent thermostatistics 
of $q$-bosons can be formulated by using an appropriate prescription of the JD 
to be used in the thermodynamic relations so that the entire structure of thermodynamics in the 
sense of the Legendre transformations is preserved. 

Many investigations are devoted to the study of 
$q$-oscillators providing much insight into both the 
mathematical development and the $q$-deformed thermodynamics 
\cite{ng,chai,lee,song,del,nar,kan,rodi,ubri}. 
However, it is our opinion that a full understanding of the physical meaning of the $q$-deformation and its effect on the thermodynamic 
relations is still lacking. 

The purpose of this paper is twofold: first, we want to extend the formalism introduced in 
Ref.\cite{pre2000} to incorporate the algebra which is  symmetric  under the transformation $q\leftrightarrow \q1$,  considering both 
boson and fermion degrees of freedom. Second, by using the generalized thermodynamic relations, 
we wish to examine the equation of state of the $q$-deformed ideal gas in the semi-classical limit and 
the low temperature regime of $q$-fermions in order to  provide a deeper insight into 
the understanding of the statistical behavior of $q$-deformed bosons and fermions. We organize the paper as follows. Sec.II introduces the $q$-deformed algebra of bosons and fermions and the thermal expectation values. The statistical thermodynamics of  $q$-deformed bosons and fermions is developed in Sec.III. The equation of state and specific heat of the ideal boson and fermion gases are derived in Sec.IV. We study the equation of state in the semi-classical limit in Sec.V.  We devote Sec. VI to the investigation of the ideal 
$q$-deformed fermions at low temperature. Sec.VII contains some concluding remarks.

\section{$q$-oscillators algebra and thermal averages}

In this section we shall briefly review the basic properties of $q$-oscillators algebra 
useful in the present  investigation. The symmetric $q$-oscillators algebra is defined 
in terms of the creation and annihilation operators 
$c$, $c^\dag$ and the $q$-number operator $N$ by \cite{ng,chai,lee,song}
\begin{equation}
[c,c]_{\kappa}=[c^\dag,c^\dag]_{\kappa}=0 \; , \ \ \ 
cc^\dag- \kappa \, q^\kappa \,  c^\dag c =q^{-N} \; , 
\end{equation}

\begin{equation}
[N,c^\dag]= c^\dag \; , \ \ \ [N,c]=-c\; ,
\end{equation}
where the  deformation parameter $q$ is real and 
$[x,\, y]_{\kappa}= x y - \kappa y x\,$,
where  $\kappa = 1$ for  $q$-bosons with commutators and $\kappa = -1$ for 
$q$-fermions with anticommutators.  

Furthermore, the operators obey  the relations
\begin{equation}
c^{\dag}c = [N]\, , \ \ \ \ c c^{\dag} = [1 + \kappa N] \, ,
\label{cc}
\end{equation}
where the $q$-basic number is defined as 
\begin{equation}
[x]=\frac{q^x-q^{-x}}{\qq1}\; .
\label{bn}
\end{equation}

By using the above algebra it is possible to construct  the $q$-Fock space 
valid for both $q$-bosons and $q$-fermions. It is,  however,  important to stress
that for $q$-fermions the eigenvalues of the number operator $N$ 
can take on the values $n=0,1$ only (as in the case of undeformed fermions).

The transformation from Fock space to the configuration space (Bargmann holomorphic representation) may be accomplished by means of the Jackson derivative (JD) \cite{jack,flo,fink}
\begin{equation}
\dqq f(x)=\frac{f(qx)-f(\q1 x)}{x\,(\qq1)}\; ,
\end{equation}
which reduces to the ordinary derivative  in the limit when $q$ goes to unity. Therefore, 
the JD occurs naturally in $q$-deformed structures and we will see that 
it plays a crucial role in the $q$-generalization of the thermodynamic relations. 

Thermal average of observables can be computed by following the usual prescription 
of quantum mechanics. The Hamiltonian of the non-interacting 
$q$-deformed oscillators (fermions or bosons) is expected to have the form
\begin{equation}
H=\sum_i (\epsilon_i-\mu) \, N_i\; ,
\label{ha}
\end{equation}
where $\mu$ is the chemical potential and $\epsilon_i$ is the kinetic energy in 
the state $i$ associated with the number operator $N_i$. 
Let us note that the Hamiltonian is deformed and depends implicitly on $q$,  since the number 
operator is deformed by means of Eq.(\ref{cc}). 
The thermal average of an operator has the standard form 
\begin{equation}
\langle {\cal O}\rangle=Tr \left (\rho\, {\cal O} \right )\; ,
\end{equation}
where $\rho$ is the density operator and $\cal Z$ is the grand canonical partition function defined as 
\begin{equation}
\rho=\frac{e^{-\beta H}}{\cal Z}\; , \ \ \ \ \ \  
{\cal Z}=Tr \left ( e^{-\beta H} \right )\; ,
\label{pf}
\end{equation}
and $\beta = 1/T$ (henceforward we shall set Boltzmann constant equal to unity).
We observe that the structure of the density matrix $\rho $ and the 
thermal average are undeformed. As a consequence, the structure of the partition 
function is also unchanged. We emphasize that this is not a trivial assumption 
because its  validity  implicitly amounts to an unmodified structure of the 
Boltzmann-Gibbs entropy, $S_q=\log W_q$,  
where $W_q$ stands for the number of states of the system corresponding 
to the set of occupation numbers $n_i$.
Obviously the number $W_q$ is modified in the $q$-deformed case.
Such a deformation is radically different from the so-called non-extensive statistics, 
recently proposed by Tsallis \cite{tsallis}, where the structure of the entropy is 
deformed via the logarithmic function.  

The above assumptions allow us to calculate 
the average occupation number $n_i$ defined by the relation 
$[n_i]= Tr \left (  e^{- \beta H} c^{\dag}_ic_i\right )/{\cal Z}\, $.
Repeated application of the algebra of $c, c^{\dag}$ along with the use 
of the cyclic property of the trace leads to the result \cite{tus}
\begin{equation}
n_{i}=\frac{1}{\qq1} 
\log\left (\frac{z^{-1}e^{\beta\epsilon_i}-\kappa \, q^{-\kappa}} 
{ z^{-1}e^{\beta\epsilon_i}-\kappa \, q^\kappa}\right) \; ,
\label{nqi}
\end{equation}
where $z = e^{\beta \mu}$ is the fugacity. 
It is easy to verify that the $q\rightarrow 1$ limit reproduces 
the standard distribution. We shall see later that such a result can be obtained 
in a natural way by means the generalization of the standard thermodynamic 
relations.

\section{Thermodynamics of $q$-deformed bosons and fermions}

In order to develop the thermodynamics of $q$-oscillators, 
we shall begin with the logarithm of the Grand partition function  given by
\begin{equation}
\log {\cal Z}=-\kappa \sum_i \log (1-\kappa \, z \, e^{-\beta\epsilon_i}) \; .
\label{part}
\end{equation}
This form is due to the fact that we have chosen the Hamiltonian to be a linear function 
of the number operator but it is  not linear in $c^\dag c$. For this reason,
the standard  thermodynamic relations in the usual form are ruled out. 
It is verified, for instance, that 
$N\ne z \, \frac{\partial}{\partial z} \log {\cal Z}$. 

As the coordinate space representation of the $q$-boson algebra is realized by the introduction of the JD, we stress  that the key point of the $q$-deformed thermostatistics is in the observation that the ordinary thermodynamic derivative with respect to $z$, must be replaced by the JD, thus
\begin{equation}
\frac{\partial}{\partial z} \Longrightarrow {\cal D}^{(q)}_z \; .
\end{equation}
%where we have defined ${\cal D}^{(q)}_z$ as the Jackson derivative up to a constant (which goes to unity when $q \rightarrow 1$)
%\begin{equation}
%{\cal D}^{(q)}_z = \frac{q-1}{\log q}\, \partial^{(q)}_z \; .
%\end{equation}
Consequently, the number of particles in the $q$-deformed theory can be derived from the relation 
\begin{equation}
N=z \; {\cal D}^{(q)}_z \log {\cal Z}\equiv \sum_i n_i \; ,
\label{num}
\end{equation}
where $n_i$ is the mean occupation number expressed in Eq.(\ref{nqi}). 

The usual Leibniz chain rule is not valid for the JD and therefore derivatives encountered in thermodynamics must be modified according to a well-defined  prescription as follows. First we observe that the JD applies only with respect to the variable in the exponential form such as $z=e^{\beta \mu}$ or $y_i=e^{-\beta \epsilon_i}$. Therefore for the $q$-deformed case, any thermodynamic derivative of functions which depend on $z$ or $y_i$ must be transformed to derivatives in one of these variables by using the ordinary chain rule and then evaluating  the JD with respect to the exponential variable.  For the  case of the internal energy in the $q$-deformed case, we can write this prescription explicitly as
\begin{equation}
U=-\left. \frac{\partial}{\partial\beta} \log {\cal Z} \right |_z=
\kappa\sum_i \frac{\partial y_i}{\partial\beta} \, 
{\cal D}^{(q)}_{y_i}\log(1-\kappa z\,y_i) \; .
\label{int}
\end{equation}
In this case we obtain the correct form of the internal energy
\begin{equation}
U=\sum_i \epsilon_i \, \nq\; ,
\label{un}
\end{equation}
where $n_i$ is the mean occupation number expressed in Eq.(\ref{nqi}). This shows that the above procedure is the correct prescription for the application of the JD.

Introducing the thermodynamic potential $\Omega = - T \log {\cal Z}$ , we can determine the entropy to be 
\begin{eqnarray}
S=-\left. \frac{\partial\Omega}{\partial T}\right |_\mu &&\equiv 
\log {\cal Z} +\kappa \, \beta\, \sum_i\left.
\frac{\partial \alpha_i}{\partial\beta}\right|_\mu  
{\cal D}^{(q)}_{\alpha_i}\log(1-\kappa \, \alpha_i)\nonumber\\
&&=\log {\cal Z} +\beta U-\beta\mu N \; ,
\label{S=}
\end{eqnarray}
where $\alpha_i=z\, e^{-\beta\epsilon_i}$,  
$N$  and $U$  are the modified functions expressed in  Eqs.(\ref{num}) 
and (\ref{int}). 
After some simple manipulations, we may express the above entropy 
in terms of the $q$-basic numbers as follows
\begin{equation}
S=\sum_i \Big \{ -\nq \, \log\, [\nq]+
\kappa (1+\kappa \nq)\,  \log\, [1+\kappa \nq]-\kappa \, 
\log \Big([1+\kappa \nq]-\kappa [\nq] \Big) \Big \} \; .
\label{entro}
\end{equation}

We point out that the entropy determined in Eq.(\ref{entro}) reduces to the standard boson and fermion entropies \cite{reichl} in the $q \rightarrow 1$ limit. 
It is interesting to observe that the above entropy has the same structure as the standard 
boson/fermion entropy apart from the appearance of the last 
term which follows from the nonadditivity property of the $q$-basic number defined in Eq.(\ref{bn}). 

%\begin{equation}
%S_1=\sum_i \Big \{ -\nq \, \log\, \nq+
%\kappa (1+\kappa \nq)\,  \log(1+\kappa\nq) \Big \}\; .
%\end{equation}

We can establish  internal self-consistency by demonstrating that the extremization of the entropy with constrained internal energy and total number of particles, leads to the correct distribution function. Accordingly, the extremization can be stated as
\begin{equation}
\delta \, \Big ( S-\beta U+\beta\mu N \Big )=0 \; ,
\label{extr}
\end{equation}
where $\beta$ and $\beta\mu$ play the role of Lagrange multipliers. Following our prescription for the use of the JD, this condition becomes
\begin{equation}
{\cal D}^{(q)}_{y_i} \, \Big ( S-\beta U+\beta\mu N \Big ) \, 
\delta y_i=0\; .
\end{equation}
After simple manipulations it is easy to show that the above condition gives 
the correct distribution function as in Eq.(\ref{nqi}) derived from the $q$-algebra.

\section{Ideal $q$-deformed boson and fermion gas}

We shall now proceed to investigate the thermodynamic functions describing the 
behavior of an ideal $q$-deformed boson or fermion gas. 
In the following, we shall not analyze the phenomena of $q$-boson condensation 
(explicitly studied in Ref.\cite{pre2000}). In other words, for $q$-bosons the 
considered temperature is greater than the critical temperature. 

For a large volume and a large number of particles, 
the sum over states can be replaced by the integral and the thermodynamic relation 
$PV/T=\log {\cal Z}$ can be written as 
\begin{equation}
\frac{P}{T}=-\kappa \frac{2}{\sqrt{\pi}} \, \frac{g_\kappa}{\lambda^3} 
\int_0^\infty \!\!\! dx \; x^{1/2} \, 
\log (1-\kappa \, z \, e^{-x})\; ,  
\label{peq}
\end{equation}
where $g_\kappa$ is the spin degeneracy factor,  
$x=\beta\epsilon$, $\epsilon = p^2/2m$ is the kinetic energy and 
$\lambda = h/(2\pi m T)^{1/2}$ is the thermal wavelength.

Following the prescription of the JD in the $q$-deformed thermodynamics derivatives, 
we may re-express the above equation as
\begin{equation}
\frac{P}{T}=\frac{g_\kappa}{\lambda^3} \; h_{_{5/2}}^\kappa (z,q) \; ,
\label{presa}
\end{equation}
where we have defined the $q$-deformed $h_{n}^\kappa (z,q)$
function as 

\begin{eqnarray}
h_n^\kappa(z,q)&=&\frac{1}{\Gamma (n)} \int_0^\infty \!\!\! dx \; x^{n-1} 
\frac{1}{\qq1} \log\left (\,\frac{z^{-1}e^x-\kappa \, q^{-\kappa}}
{ z^{-1}e^x-\kappa \, q^\kappa} \right) \nonumber\\
&\equiv& \frac{1}{\qq1} \left ( 
\sum_{i=1}^{\infty} \frac{(\kappa \, q^\kappa \, z)^i}{i^{n+1}} - 
\sum_{i=1}^{\infty} \frac{(\kappa \, q^{-\kappa} \, z)^i}{i^{n+1}}   
\right )  \; .
\label{hn}
\end{eqnarray} 

In the limit $q\rightarrow 1$, the deformed $h_n^\kappa(z,q)$ functions 
reduce to the standard  $g_n(z)$ functions for bosons and to the $f_n(z)$ functions 
for fermions \cite{reichl}. 
In terms of these $q$-generalized functions, we obtain the particle  density,
\begin{equation}
\frac{N}{V}=\frac{g_\kappa}{\lambda^3} \; h_{_{3/2}}^\kappa (z,q) \; ,
\label{totnpa}
\end{equation}
and the internal energy, 
\begin{equation}
U=\frac{3}{2} \; \frac{g_\kappa}{\lambda^3} \; VT 
\; h_{_{5/2}}^\kappa (z,q) \; .
\label{ueq}
\end{equation}
Comparing Eqs.(\ref{presa}) and (\ref{ueq}), 
we see that, as in the case of the undeformed gas, the following well-known relation is satisfied 
\begin{equation}
U=\frac{3}{2} \, PV \; .
\label{eos}
\end{equation}
In  a similar manner, in the thermodynamic limit, we can obtain the entropy per unit volume 
\begin{equation}
\frac{S}{V}=\frac{g_\kappa}{\lambda^3} 
\left(\,\frac{5}{2} \; h_{_{5/2}}^\kappa (z,q)- 
h_{_{3/2}}^\kappa (z,q) \log z \right) \; .
\label{entroa}
\end{equation}

We may now proceed to calculate the specific heat
\begin{equation}
C_v=\left. \frac{\partial U}{\partial T}\right|_{V,N}\; .
\label{cvt}
\end{equation}
Making use of the JD prescription as before, we determine the specific heat to be
\begin{equation}
C_v=-\beta^2 \sum_i \, \epsilon_i\,\frac{\partial \alpha_i}{\partial \beta}\; 
\frac{1}{\qq1} {\cal D}_{\alpha_i}^{(q)}\log 
\left ( \frac{1- \kappa \, q^{-\kappa} \alpha_i}{1-\kappa \, q^\kappa \alpha_i}  
\right ) \; ,
\end{equation}
where $\alpha_i=z e^{-\beta\epsilon_i}$. The above equation can be written in the thermodynamic 
limit as 
\begin{equation}
C_v= \frac{g_\kappa}{\lambda^3} \Big \{ \frac{15}{4}\,
z\, {\cal D}^{(q)}_z h_{_{7/2}}^\kappa (z,q)
-\frac{9}{4}\; 
\frac{z\; [{\cal D}^{(q)}_z h_{_{5/2}}^\kappa (z,q)]^2}
{ {\cal D}^{(q)}_z h_{_{3/2}}^\kappa (z,q)} \Big \} \; .
\label{cva}
\end{equation}
We observe that the above equations have the same structure as the 
undeformed relation, even though the deformation is contained in the JD and in the 
$h (z,q)$ functions.

\section{Equation of state in the semi-classical limit}

First, we observe that in the classical limit $z=e^{\beta \mu} \ll 1$,  
the $q$-deformed distribution function (\ref{nqi}) 
reduces to the standard 
Maxwell-Boltzmann distribution and the entropy (\ref{entro}) reduces to the 
Boltzmann entropy: $S=-\sum_i n_i \log n_i$, for any value of $q$. 
Furthermore, studying the classical limit in Eq.(\ref{cva}), it is easy to check that the 
specific heat goes to its classical value: $C_v \rightarrow 3/2 N$.
Hence the $q$-deformation in the thermodynamic relations is a pure 
quantum effect, which is washed up in the classical limit.  
Such a feature is quite the opposite in the case of Tsallis statistics where the deformation is 
intrinsic and does not depend on the quantum nature of the system \cite{tsallis}.

We shall now  investigate the effects of the deformation in  the 
equation of state of an ideal quantum boson/fermion gas in the 
semiclassical limit.  
For low values of $z$, Eq.(\ref{peq}) reduces to 
\begin{equation}
P \simeq\frac{T}{\lambda^3} \, g_\kappa \, z \, 
\Big \{1+\kappa \, \Big (\frac{q+\q1}{2}\Big) \, \frac{z}{2^{5/2}} \Big\} \; .
\label{pveq}
\end{equation}
The equation of state can be expressed in terms of the 
number of particles considering the semiclassical limit of Eq.(\ref{totnpa})
\begin{equation}
N\simeq\frac{V}{\lambda^3} \, g_\kappa \, z \, 
\Big \{1+\kappa \, \Big (\frac{q+\q1}{2}\Big ) \, \frac{z}{2^{3/2}} \Big \}\; .
\end{equation}
Inverting the above equation as $z=z(N)$ and inserting the result in Eq.(\ref{pveq}), 
we can write the equation of state in the semiclassical limit as 
\begin{equation}
PV\simeq NT \Big [ 1- \kappa \, \Big (\frac{q+\q1}{2}\Big ) \, 
\frac{ N \lambda^3}{V \,  g_{\kappa} \, 2^{5/2}} \Big \}\, .
\label{pvnt}
\end{equation} 

As usual, when the thermal wavelength $\lambda$ is much less than the average 
inter-particle distance ($\lambda^3 \ll V$), the quantum statistics does not have a 
significant influence on the thermodynamic property of the gas. 
Otherwise we find that the equation of state is modified by the quantum statistical 
effect and, at fixed volume, the pressure is decreased for $q$-boson particles 
($\kappa=1$) and increased for $q$-fermion particles ($\kappa=-1$) compared
to the classical case. 
This feature is similar to the standard boson/fermion result: the attractive boson 
interactions reduce the pressure and the repulsive fermion interactions increase the 
degeneracy pressure. However, in Eq.(\ref{pvnt}), 
this effect is enhanced by the factor $(q+\q1)/2$ (always greater than unity). 
Therefore, the $q$-deformation of the algebra implies an enhancement of the  
quantum statistical behavior of the particles relative to the standard bosons or fermions. 
In the next section, we will see that this feature is in agreement with the results 
obtained considering $q$-deformed fermions at low temperature regime.

\section{Ideal $q$-deformed fermions at low temperature}

We shall now  investigate the low temperature behavior of the 
$q$-deformed fermions ($\kappa=-1$). 
We start by observing that for the case of  $q$-fermions, the Fermi surface 
in the limit $T\rightarrow 0$ 
(or $\beta\rightarrow +\infty$)
assumes the standard undeformed step shape, for all values of $q$: 
$n_q \rightarrow \vartheta (\mu-\epsilon)$. 
Consequently, the quantum deformation of the statistics is a finite temperature effect only. 

In order to bring out the salient features of the behavior of the $q$-deformed fermion gas at 
low but finite temperature, let us begin with the function introduced in Eq.(\ref{hn}), 
now with $\kappa =-1$ which is needed to evaluate the number of particles for unit volume 
and the internal energy. 
Thus we specifically consider the integral
\begin{eqnarray}
f_{_{3/2}}(z,q)=\frac{2}{\sqrt{\pi}} \int_0^\infty \!\!\! dx \; x^{1/2} 
\frac{1}{\qq1} \log\left (\,\frac{e^{x-\nu}+ \, q}
{e^{x-\nu}+ \, q^{-1}} \right)\; ,
\end{eqnarray}
where $\nu=\beta\mu$. 
Considering the low temperature limit $\nu\rightarrow\infty$, 
the above integral can be expanded in a Taylor series as follows 
\begin{eqnarray}
f_{_{3/2}}(z,q)=\frac{2}{\sqrt{\pi}}\Big \{ 
\frac{2}{3}\, \nu^{3/2}+\gamma_1(q) \frac{\pi^2}{12} \, \nu^{-1/2}+
\gamma_3(q) \frac{7\pi^4}{120}\, \frac{\nu^{-5/2}}{8} +  \cdots \Big \} \; ,
\label{f32}
\end{eqnarray}
where $\gamma_n(q)$ is a factor which depends on the deformation parameter 
\begin{eqnarray}
\gamma_n(q)=
\int_0^\infty \!\!\! dt \; t^n \;  
\frac{1}{\qq1} \log\left (\,\frac{e^{t}+ \, q}
{e^{t}+ \, q^{-1}} \right) \Big /
\int_0^\infty \!\!\! dt \; t^n \;  \frac{1}{e^t+1} \; , 
\end{eqnarray}
and goes to unity when $q\rightarrow 1$. 

Inserting Eq.(\ref{f32}) in  Eq.(\ref{totnpa}), we have
\begin{eqnarray}
\frac{1}{g_F}\frac{N}{V} \, \left ( \frac{2\pi\hbar^2}{m}\right )
=\frac{4}{3 \sqrt{\pi}}\Big \{ 
\nu^{3/2}+\gamma_1(q) \frac{\pi^2}{8} \, \mu^{-1/2}\, T^2+
\gamma_3(q) \frac{7\pi^4}{640}\, \nu^{-5/2} \, T^4 + \cdots  \Big \} \; .
\label{nveq}
\end{eqnarray}
In the limit $T\rightarrow 0$, we obtain  the  expression for the chemical potential (equivalent to the 
Fermi energy) as a function of the average density  in the form
\begin{equation}
\mu(T=0)\equiv\epsilon_F=\frac{\hbar^2}{2m}\, 
\left (\frac{6 \pi^2}{g_F}\frac{N}{V}\right )^{2/3} \; .
\label{mutzero}
\end{equation}
Since the deformation acts only at finite temperature,  the Fermi energy 
does not depend on $q$.

If we now consider the effect to second order in T, the chemical potential can be written as 
\begin{equation}
\mu=\epsilon_F \left [ 1-\gamma_1(q)\, \frac{\pi^2}{12}\,  
\left ( \frac{T}{\epsilon_F}\right )^2 \right ] \; .
\end{equation} 
Because $\gamma_1(q)<1$ for $q\ne 1$ and decreases when  the deformation parameter is increased
(for example, $\gamma_1(1.5)=0.989$, 
$\gamma_1(2)=0.969$), the deformed quantum statistics 
implies a reduction of the chemical potential at finite temperature but this 
reduction is smaller than the undeformed fermion case. 
In other words, the $q$-deformation increases
the strength of the Pauli repulsion between particles at finite temperature in clear agreement with the result obtained in the previous section. A somewhat similar result was found by Greenberg 
in the framework of quon statistics \cite{green}. 

Let us now  compute the internal energy,  which in the low temperature regime becomes 
\begin{eqnarray}
\frac{U}{V}=\frac{3}{2}\, \frac{g_F}{\lambda^3} \, T \, f_{5/2}(z,q)\simeq 
&&\frac{4}{5\sqrt{\pi}}\, g_F \, \left ( \frac{m}{2\pi\hbar^2}\right )^{3/2}  
\Big [ \mu^{5/2}+\gamma_1(q) \frac{5 \pi^2}{8}  \, 
\mu^{1/2} \, T^2 \nonumber \\
&&-\gamma_3(q) \frac{7\pi^4}{384}\, \mu^{-3/2} \, T^4 
+ \cdots   \Big ] \; .
\end{eqnarray}
By using Eqs.(\ref{nveq}) and (\ref{mutzero}), the above expression 
at the second order in $T$ can be written as

\begin{equation}
U=\frac{3}{5} \, N \, \epsilon_F 
\left [ 1 +\gamma_1(q) \frac{5 \pi^2}{12}\, 
\left (\frac{T}{\epsilon_F}\right )^2 \right ]\; .
\end{equation}

Similarly, the heat capacity in the low temperature regime and for $q\approx 1$ can be obtained as 
\begin{equation}
C_v=\left. \frac{\partial U}{\partial T}\right|_{V,N}=
\gamma_1(q) \, 
\frac{N \, \pi^2}{2 \, \epsilon_F} \,  \; T \; .
\label{cv}
\end{equation}
The $q$-fermionic specific heat depends linearly on $T$ at very 
low temperatures and goes to zero at zero temperature in accordance with the 
third law of thermodynamics. In Eq.(\ref{cv}) the specific heat is modulated by the 
factor $\gamma_1(q)$ which is always smaller than unity for $q\ne 1$, therefore the 
specific heat at low temperature is smaller in $q$-deformed theories as a consequence of 
a greater Pauli repulsion at non-zero temperature.

\section{Conclusion}
We have studied the thermostatistics of $q$-bosons and $q$-fermions using 
the symmetric $q\leftrightarrow \q1$ algebra and we have shown that the entire
structure of thermodynamics is preserved if the ordinary derivatives are replaced 
by the JD. This prescription gives us the recipe to derive the fundamental 
thermodynamic functions and an explicit form of the $q$-deformed entropy. 

In the classical limit the $q$-deformed statistics is reduced to the 
classical statistical mechanics: the distribution functions goes to the Maxwell-Boltzmann distribution 
and the entropy reduces to the Boltzmann entropy. The $q$-deformation is a pure quantum effect.
However, we have seen that $q$-deformed bosons and fermions have an enhancement 
of the quantum statistical effects compared to standard behavior: at finite temperature, the 
ideal gas of $q$-bosons is more attractive than undeformed ideal boson gas, and 
for $q$-fermions, the Pauli repulsion is stronger than undeformed 
fermions. This feature is confirmed by an examination of  the low temperature limit of $q$-deformed 
fermion ideal gas. At zero temperature the Pauli exclusion principle is rigorously satisfied  
but at low but non-zero temperature,  the chemical potential is greater than in the undeformed case 
and the specific heat is smaller. Such an effect appears more significant  by increasing the 
deformation parameter $q$. 

The generalized thermostatistics which we have developed appears to provide a deeper insight into the 
nature of the deformed boson and fermion algebra. These results can be conceptually important 
in many physical fields from solid state to cosmological problems. For example, primordial 
nucleosynthesis can be non-trivially modified by the influence of statistics \cite{torres}. 
Furthermore, finite limits on Pauli principle violation by nuclei produced in the core 
collapse supernovas has been found in Ref.\cite{baron}. Since  supernova explosions are   finite temperature 
events ($T \simeq 430$ keV), a value of the $q$-deformation parameter slightly different from unity 
could take into account such effects and influence sensibly 
the stellar collapse mechanism.

\end{document}